\documentclass[letterpaper,twoside,journal,final,twosided]{IEEEtran}
\usepackage[T1]{fontenc}
\usepackage[latin9]{inputenc}
\usepackage{amsmath}
\usepackage{amssymb}
\usepackage{graphicx}
\usepackage[unicode=true,pdfusetitle,
 bookmarks=true,bookmarksnumbered=false,bookmarksopen=false,
 breaklinks=false,pdfborder={0 0 1},backref=false,colorlinks=false]
 {hyperref}
\usepackage{breakurl}

\makeatletter

\newcommand{\lyxdot}{.}

\setkeys{Gin}{width=\linewidth}
\usepackage{cite}

\long\def\@makecaption#1#2{\ifx\@captype\@IEEEtablestring%
\footnotesize\begin{center}{\normalfont\footnotesize #1}\\
{\normalfont\footnotesize\scshape #2}\end{center}%
\@IEEEtablecaptionsepspace
\else
\@IEEEfigurecaptionsepspace
\setbox\@tempboxa\hbox{\normalfont\footnotesize {#1.}~~ #2}%
\ifdim \wd\@tempboxa >\hsize%
\setbox\@tempboxa\hbox{\normalfont\footnotesize {#1.}~~ }%
\parbox[t]{\hsize}{\normalfont\footnotesize \noindent\unhbox\@tempboxa#2}%
\else
\hbox to\hsize{\normalfont\footnotesize\hfil\box\@tempboxa\hfil}\fi\fi}
\ifCLASSOPTIONcompsoc
  \else
  \fi

\usepackage{dblfloatfix}

\@ifundefined{showcaptionsetup}{}{%
 \PassOptionsToPackage{caption=false}{subfig}}
\usepackage{subfig}
\makeatother

\begin{document}

\title{Time-Frequency Warped Waveforms}

\author{Mostafa Ibrahim, Ali Fatih Demir, Student Member, IEEE, Hüseyin Arslan,
Fellow, IEEE\vspace{-0mm}
\thanks{Manuscript received September 15, 2018; revised November
04, 2018; and accepted November 07, 2018.} \thanks{Mostafa Ibrahim
is with the College of Engineering, Istanbul Medipol University, Istanbul,
Turkey (e-mail: mostafa.ibrahim.kamel@gmail.com).}\thanks{Ali Fatih
Demir is with the Department of Electrical Engineering, University
of South Florida, Tampa, FL, USA (e-mail: afdemir@mail.usf.edu).}\thanks{Hüseyin
Arslan is with the Department of Electrical Engineering, University
of South Florida, Tampa, FL, USA and also with the College of Engineering,
Istanbul Medipol University, Istanbul, Turkey (e-mail: arslan@usf.edu).}}
\maketitle
\begin{abstract}
The forthcoming communication systems are advancing towards improved
flexibility in various aspects. Improved flexibility is crucial to
cater diverse service requirements. This letter proposes a novel waveform
design scheme that exploits axis warping to enable peaceful coexistence
of different pulse shapes. A warping transform manipulates the lattice
samples non-uniformly and provides flexibility to handle the time-frequency
occupancy of a signal. The proposed approach enables the utilization
of flexible pulse shapes in a quasi-orthogonal manner and increases
the spectral efficiency. In addition, the rectangular resource block
structure, which assists an efficient resource allocation, is preserved
with the warped waveform design as well. 
\end{abstract}


\markboth{IEEE COMMUNICATIONS LETTERS}{Ibrahim \MakeLowercase{\textit{et al.}}: Time-Frequency Warped Waveforms}

\section{Introduction}

\IEEEPARstart{T}{he} forthcoming wireless communication technologies
are envisioned to support a vast variety of services. Recently, the
International Telecommunications Union (ITU) has characterized the
use cases for the fifth generation (5G) mobile networks as enhanced
mobile broadband (eMBB), massive machine type communications (mMTC),
and ultra-reliable low-latency communications (URLLC) featuring 20
Gb/s peak data rate, 10\textsuperscript{6}/km\textsuperscript{2}
device density, and less than 1 ms latency respectively \cite{zhang2016}.
Therefore, a flexible design is needed to satisfy these challenging
requirements and, the waveform, which is the essential component of
an air interface, has to be designed precisely to accommodate such
flexibility.

Orthogonal frequency-division multiplexing (OFDM) and its low peak-to-average-power
ratio (PAPR) variant, discrete Fourier transform spread OFDM (DFT-s-OFDM)
are the most favored waveforms that are being implemented in various
standards. They provide several attractive attributes such as effective
hardware implementation, low-complexity equalization, and straightforward
multiple-input-multiple-output (MIMO) integration. In addition, the
numerology concept (i.e., adaptive waveform parametrization) in the
upcoming 5G standard \cite{parkvall2017} will improve the flexibility
of these waveforms. On the other hand, OFDM and DFT-s-OFDM have severe
high out-of-band emissions (OOBE) issue, which causes interference.
Also, any disparity in waveform parametrization (such as subcarrier
spacing) causes loss of orthogonality and leads to interference as
well \cite{ankarali2017}. Commonly, the interference amount is managed
by performing various windowing/filtering operations along with the
guard allocation \cite{demir2017}. Numerous waveforms have been proposed
to provide better time-frequency concentration while providing satisfactory
spectral efficiency and sufficient flexibility \cite{zhang2016,demir2018a}.
To fully exploit and further enhance the potential of the flexible
air interface, it is necessary to increase the flexibility in waveform
design.

This paper proposes a novel waveform design scheme that utilizes flexible
pulse shapes and exploits axis warping transform \cite{baraniuk1993,baraniuk1995}
to maintain the orthogonality. More specifically, the warping transform
is non-uniform manipulation of the lattice samples and provides flexibility
to handle the time-frequency occupancy of a signal. The utilization
of flexible pulse shapes has been thoroughly investigated with certain
trade-offs. For example, it is known that the edge subcarriers of
an OFDM signal have a significant contribution to the OOBE. Therefore,
raised cosine (RC) windows with higher sidelobe suppression capabilities
are applied to edge subcarriers while lighter windows are applied
to inner ones \cite{sahin2011a}. However, this approach shortens
the cyclic prefix (CP) duration that is designated for the multipath
channel and may cause inter-symbol-interference (ISI). In another
study \cite{ankarali2014}, RC filters of a filter bank multi-carrier
(FBMC) signal are utilized with different roll-off factors to provide
flexibility. Nonetheless, this technique does not retain the orthogonality
of the pulses and leads to inter-carrier-interference (ICI). Despite
the fact that the ISI and ICI problems in these studies can be mitigated
with multi-user diversity, it is not always guaranteed. The proposed
approach in this paper enables the utilization of flexible pulse shapes
in a quasi-orthogonal manner and increases the spectral efficiency.
In addition, the rectangular resource block structure, which assists
an efficient resource allocation, is preserved with the warped waveform
design as well.

The rest of the paper is organized as follows. Section \ref{sec:II}
provides the foundations of axis warping transform. Section \ref{sec:III}
demonstrates an implementation of the warped waveform concept and
provides various performance evaluations. Section \ref{sec:IV} summarizes
the contributions and concludes the paper.

\section{Axis Warping Transformation\label{sec:II}}

\begin{figure}[b]
\centering\subfloat[]{\centering\includegraphics[width=0.5\columnwidth]{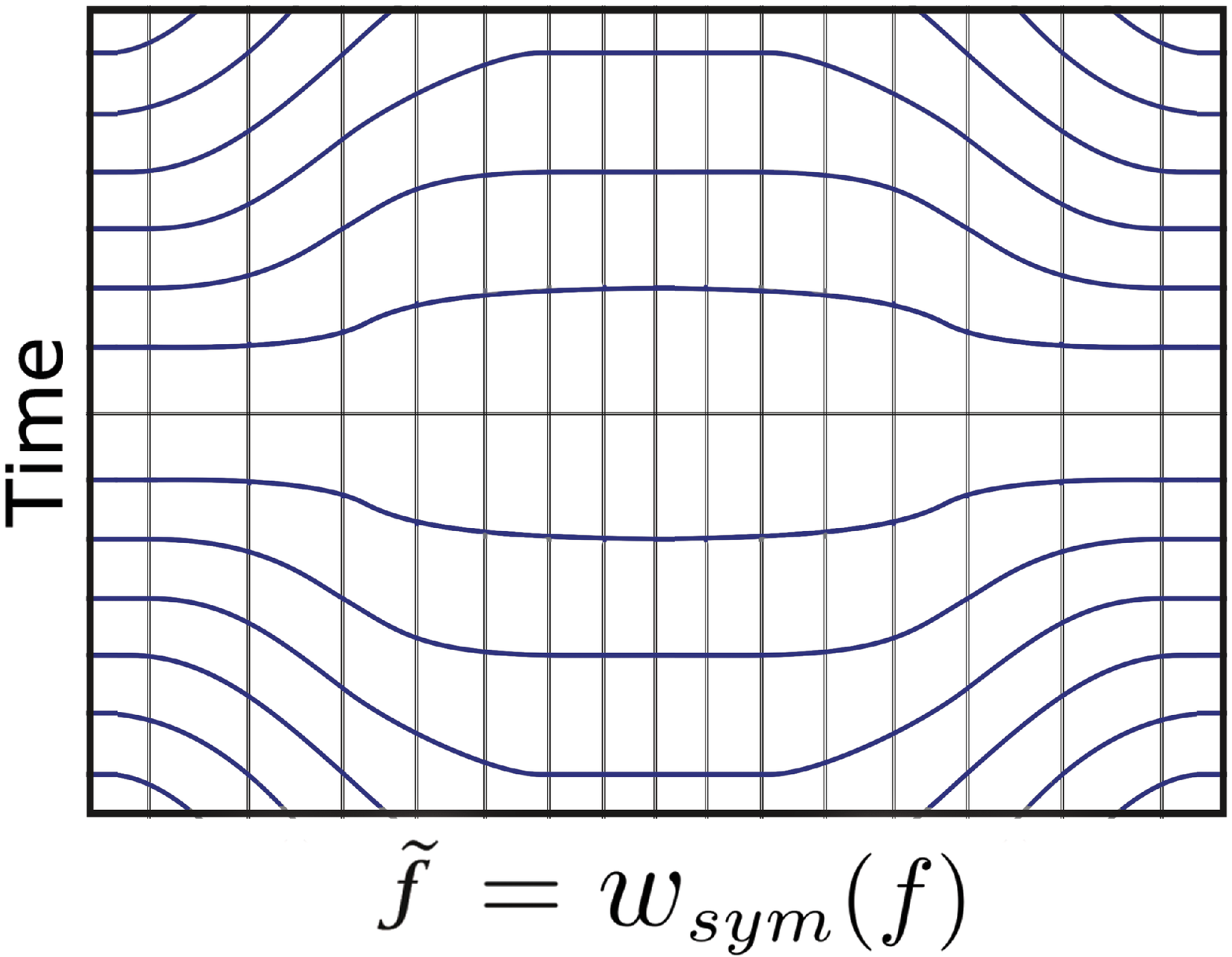}

}\centering\subfloat[]{\centering\includegraphics[width=0.5\columnwidth]{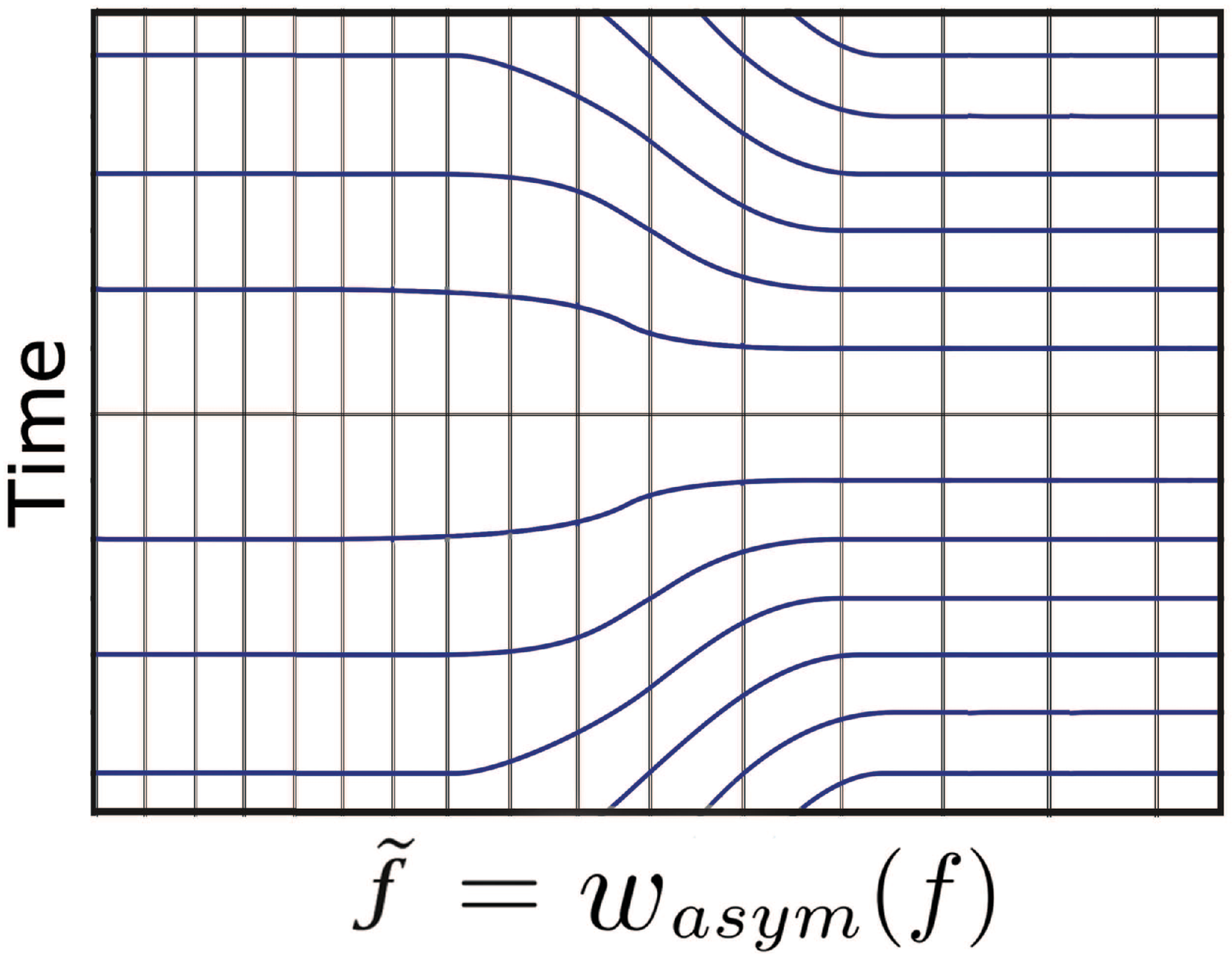}

}

\caption{Warped time-frequency plane: (a) Symmetric warping operator; (b) Asymmetric
warping operator. \label{fig:T-F Warping}}
\end{figure}

A unitary operator $\mathbf{U}$ is a linear transformation that maps
a Hilbert space onto another one (i.e., $\mathbf{U}:\mathbf{L}^{2}(\mathbb{R})\longmapsto\mathbf{L}^{2}(\mathbb{R})$).
Unitary operators preserves energy (i.e.,$\parallel\mathbf{U}s\parallel^{2}=\parallel s\parallel^{2}$),
and inner product ($\langle\mathbf{U}s,\mathbf{U}h\rangle=\langle s,h\rangle$).
Axis warping is a subclass of unitary transformations and can be expressed
as follows \cite{baraniuk1995}:

\begin{equation}
(\mathbf{U}s)(x)=\vert\dot{w}(x)\vert^{1/2}s[w(x)]\label{eq:Warping}
\end{equation}
where $w$ is a smooth, one-to-one function that comprises a large
subclass of unitary transformations. For example, a hyperbolic tangent
function ($tanh(x)$) or a sigmoid function ($sig(x)$) can serve
as a warping operator. Also, the term $\dot{w}$ represents the first
derivative of the function $w$. The relationship between the original
and warped axis is formulated by the following equation:

\begin{equation}
\tilde{x}=w(x)~~,~~\tilde{y}=y~\dot{m}(w(x))
\end{equation}
where $m=w^{-1}$ represents the inverse function of $w$, $x$ and
$y$ stand for orthogonal axis such as time and frequency, and $\tilde{x}$
and $\tilde{y}$ are the warped axes. The warping transform enables
non-uniform manipulation of axes and provides a flexible method to
control the time-frequency occupancy of the signal. Fig. \ref{fig:T-F Warping}
presents two exemplary frequency-axis warping scenarios considering
a symmetric and an asymmetric warping operators as follows:

\begin{equation}
\dot{w}_{sym}(f)=(1-\frac{1}{2}sig(\frac{|f|-a}{b}))\label{eq:Warp_Sym}
\end{equation}

\begin{equation}
\dot{w}_{asym}(f)=\big(1-\frac{1}{2}sig(\frac{f-a}{b})\big)\label{eq:Warp_Asym}
\end{equation}
where $sig(x)$ is an ``S''-shaped monotonic function that have
a range of $[0,1]$, and $a$ and $b$ represent scaling parameters.
Therefore, the given warping operators smoothly warp the frequency
axis from $f$ to twice its scale $2f$. The choice of symmetricity
depends on the application and will be detailed in the following sections.

\begin{figure}[t]
\subfloat[Subcarrier specific roll-off factors ($\alpha_{n}$) of the warped
multicarrier scheme. \label{fig:RoF}]{\centering\includegraphics[width=0.9\columnwidth]{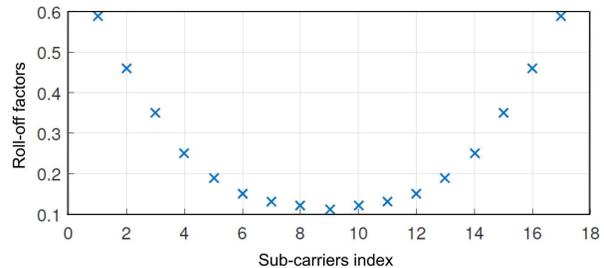}

}

\subfloat[The warped multicarrier scheme in the frequency domain. \label{fig:Warped-subcarriers}]{\centering\includegraphics[width=0.9\columnwidth]{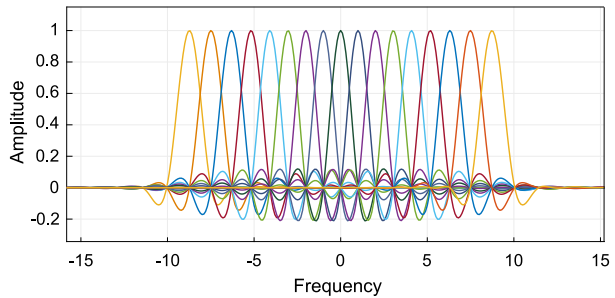}

}

\subfloat[The warped multicarrier scheme in the time domain. \label{fig:Warped-windows}]{\centering\includegraphics[width=0.9\columnwidth]{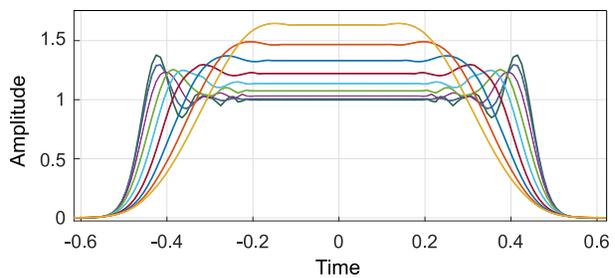}

}

\caption{A warped multicarrier scheme with $N=17$. \label{fig:Warped_MC}}
\end{figure}

\begin{figure*}[b]
\centering\subfloat[]{\centering\includegraphics[width=0.99\columnwidth]{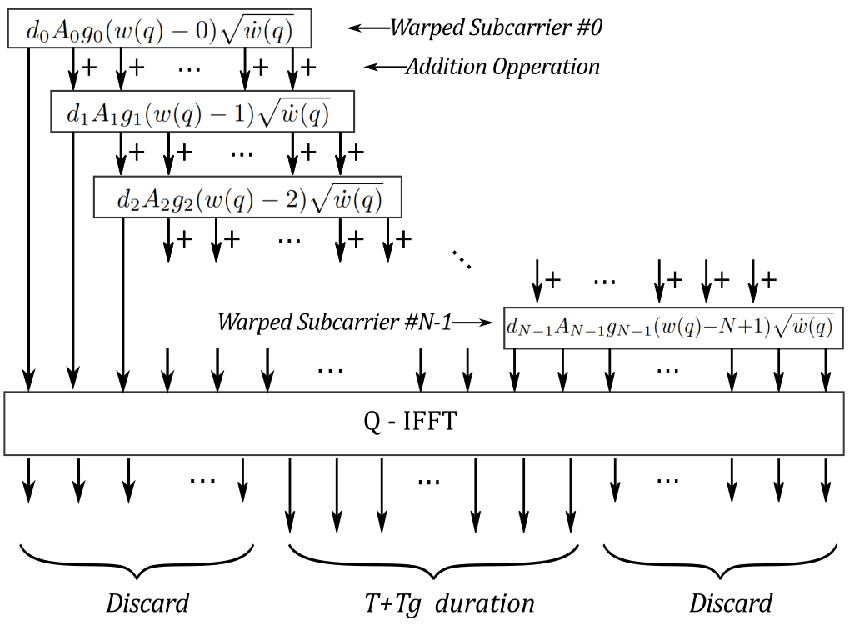}

}\centering\subfloat[]{\centering\includegraphics[width=0.99\columnwidth]{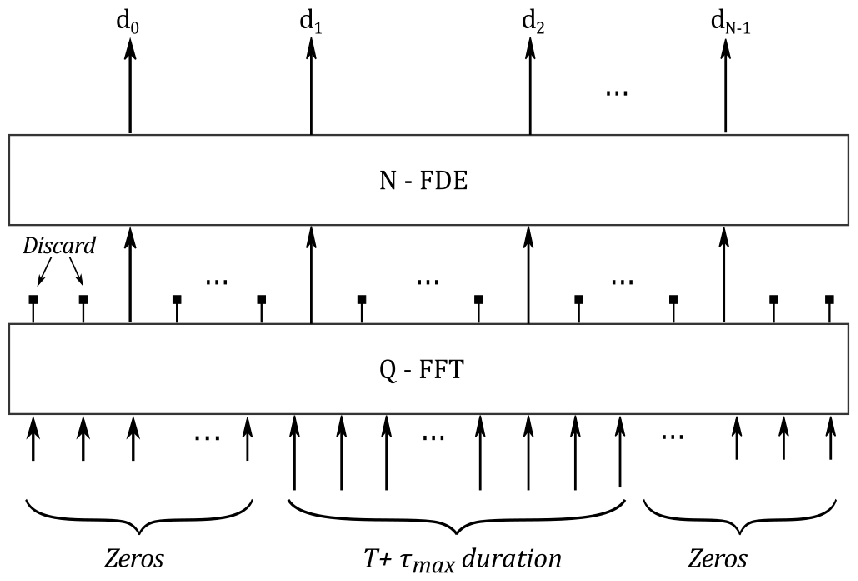}

}

\caption{Transceiver design: (a) Transmitter; (b) Receiver. \label{fig:trx}}
\end{figure*}

\section{Frequency-Axis Warping Implementation\label{sec:III}}

\subsection{Transceiver Design\label{subsec:I-TR_design}}

The warped waveform concept is demonstrated with a multicarrier scheme
that utilizes flexible pulse shaping filters. The pulse shapes that
have lower sidelobes are assigned only to the edge subcarriers in
order to decrease the OOBE in a spectrally efficient way. Afterward,
axis warping is performed to maintain orthogonality between different
pulse shapes and ensure that all subcarriers occupy the same time
window. The baseband warped waveform with $N$ subcarriers can be
expressed as follows: 
\begin{equation}
s(t)=\sum_{n=0}^{N-1}\mathbb{F}^{-1}\left[d_{n}A_{n}g_{n}(w(f)-n)\sqrt{\dot{w}(f)}\right]\label{eq:Warped_MC_cont}
\end{equation}
where $\mathbb{F}^{-1}$ represents the inverse Fourier transform,
$d_{n}$ stands for the complex data symbol, $A_{n}$ is the amplitude
equalization factor, and $g_{n}$ is the pulse shaping filter. The
RC pulse shapes with different roll-off factors ($\alpha$) are utilized
in this study. However, the choice of pulse shaping filter is not
limited to RC, and any apodization function is suitable as long as
Nyquist criteria are valid. Therefore, the design has two main variables
to shape the signal, namely $\alpha$ and $w(f)$. Also, there is
a constraint on the design to preserve the rectangular resource block
structure: all warped subcarriers should occupy the same duration
$T$. Nonetheless, the warping transform maps the complex Fourier-domain
sinusoid $e^{j2\pi f}$ to a chirp function $\sqrt{\dot{w}(f)}e^{j2\pi w(f)}$.
This mapping is realized as a chirp convolution with the RC window
in the time domain. Due to the spectral nature of nonlinear chirps
\cite{cook2016}, the convolution results in time dispersion. Accordingly,
the minimum allowed power ratio ($\zeta$) of a subcarrier within
the duration $T$ is set as a design criterion based on the performance
requirements. The relationship between the subcarrier specific roll-off
factor ($\alpha_{n}$) and $w(f)$ can be formulated considering $\zeta$
constraint as follows: 
\begin{equation}
\alpha_{n}=\underset{\alpha_{n}}{\arg\min}\Bigg\vert\frac{\int_{T}\left(\mathbb{F}^{-1}\left[R_{\alpha_{n}}(w(f)-n)\sqrt{\dot{w}(f)}~\right]\right)^{2}dt}{\int_{\mathbb{R}}\left(\mathbb{F}^{-1}\left[R_{\alpha_{n}}(w(f)-n)\sqrt{\dot{w}(f)}~\right]\right)^{2}dt}-\zeta\Bigg\vert\label{RoF}
\end{equation}
The subcarriers are orthogonal to each other only if $\zeta=100\%$.
However, $\zeta$ can be slightly lower than the ideal case in practice,
and they are considered to be quasi-orthogonal.

Fig. \ref{fig:Warped_MC} presents the warped multicarrier concept
with low number of subcarriers ($N=17$) for a better illustration.
A symmetric warping operator is selected as $w(f)=4~tanh(f/8)+0.5f$.
The subcarrier specific roll-off factors ($\alpha_{n}$) are calculated
using Eq. \ref{RoF} considering $\zeta=99.9\%$ for $T=1.1$. As
$\alpha$ increases, the frequency domain localization of an RC pulse
increases but it extends in the time domain. Though, $w(f)$ warps
the frequency axis and provides equal time domain occupation for RC
pulses with different $\alpha$s. Since the chirp function is convolved
with the RC windows in the time domain, dispersion occurs as shown
in Fig. \ref{fig:Warped-windows}. Also, please note that the subcarrier
spacing increases on the edges as a result of non-uniform manipulation
of $w(f)$, and $A_{n}$ compensates the amplitude decrease due to
the $\sqrt{\dot{w}(f)}$ factor in Eq. \ref{eq:Warping}.

A block diagram of the proposed transceiver design is shown in Fig.
\ref{fig:trx}, and the discrete form of Eq. \ref{eq:Warped_MC_cont}
is expressed for better understanding as follows: 
\begin{equation}
s\left[k\right]=\frac{1}{Q}\sum_{n=0}^{N-1}\sum_{q=0}^{Q-1}d_{n}A_{n}g_{n}(w[q]-n)\sqrt{\dot{w}[q]}e^{j2\pi kq/Q}\label{eq:Warped_MC_disc}
\end{equation}
where $k$ and $q$ are the discrete time and frequency indices. $Q$
represents the size of the FFT/IFFT operation and $Q>N$ due to the
warping expansion as mentioned earlier. Nonuniform fast Fourier transform
(NUFFT) is required at the receiver to deal with the warped time-frequency
lattice. A low complexity NUFFT implementation is proposed by coupling
the FFT operation with an interpolation scheme that interpolates to
the neighboring equispaced points \cite{fessler2003}. However, this
method is prone to the interpolation errors that cause ICI, and the
following is proposed in our implementation: 
\begin{enumerate}
\item Using high order FFT/IFFT to have a better resolution in the frequency
domain. $Q$ is an integer multiple of $N$ and hence $g_{n}$ is
upsampled. 
\item Designing the discrete $w(f)$ (i.e., $w[g]$) in such a way that
the subcarrier positions meet one of the transform branches. Hence,
$(w[q]-n)$ should be an integer. 
\end{enumerate}
In the presence of a time dispersive channel, the signal reaches to
the receiver with a spread over period $T+\tau_{max}$, where $\tau_{max}$
is the maximum delay spread of the channel. The transmitter leaves
a guard duration ($T_{g}$), which is longer than $\tau_{max}$ to
ensure the circularity of effective channel. At the receiver, the
$T+\tau_{max}$ duration is padded with zeros, and a $Q$-size FFT
is used to obtain the frequency domain representation. Afterward,
the signal is downsampled properly, by discarding the oversampling
introduced by the transmitter. A simple frequency domain equalizer
(FDE) is performed to handle the multipath channel effect as similar
to conventional OFDM systems.

\subsection{Performance Evaluation \label{subsec:III-Performance}}

The performance of the warped waveform scheme is evaluated and compared
with the conventional windowed-OFDM (W-OFDM) where all subcarriers
are treated same for sidelobe suppression. The numerical evaluations
are performed considering $N=122$ ($6$ subcarriers are left empty
due to the warping expansion) with an upsampling ratio of $u=4$ (hence
$Q=512$). A symmetric warping operator is used and the discrete warping
function is designed in such a way that it has a decreasing slope
profile over the $7$ edge subcarriers. More specifically, $\dot{w}[q]$
changes from $1/8$ to $1/4$ whereas subcarrier specific $\alpha$
changes from $1$ to $0.05$ at the edges. The inner subcarriers have
a fixed $\alpha=0.03$, since they don't contribute to the OOBE significantly.
To make a fair comparison, the warped waveform performance is compared
with the performance of W-OFDM symbol that has $N=122$ ($6$ subcarriers
are left as guards) and $\alpha=0.03$ (i.e., fixed for all subcarriers).
Therefore, both symbols have the same spectral efficiency in terms
of $bits/s/Hz$.

\begin{figure}
\centering\includegraphics[width=0.9\columnwidth]{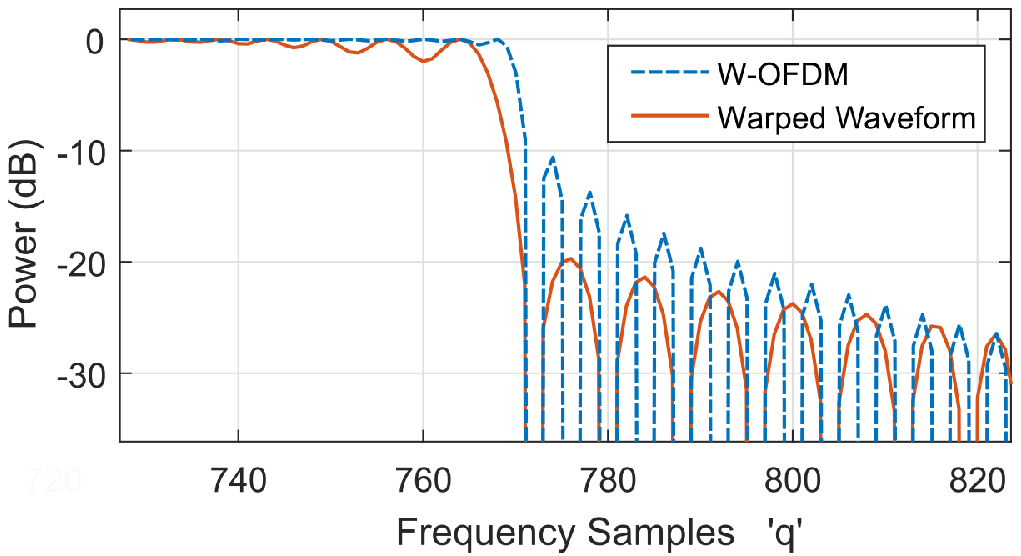}\caption{OOBE comparison of W-OFDM and warped waveform. \label{fig:OOBE}}
\end{figure}

\begin{figure}[b]
\centering\includegraphics[width=0.9\columnwidth]{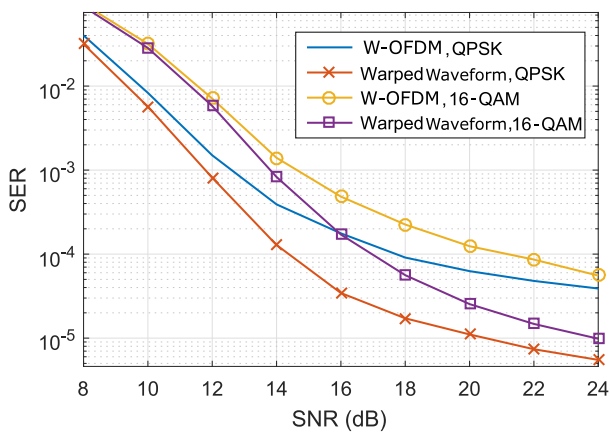}\caption{SER comparison of W-OFDM and warped waveform. \label{fig:SER}}
\end{figure}

Fig. \ref{fig:OOBE} shows that the OOBE of the warped waveform scheme
is lower than the OOBE of W-OFDM. In other words, the warped waveform
provides OOBE suppression in a more spectrally efficient way compared
to W-OFDM that requires higher $\alpha$ to achieve the same level.
Also, the symbol error rate (SER) performance is evaluated in the
presence of adjacent channel interference. The performance of the
waveforms are compared for the same spectral efficiency with the assumptions
of additive white Gaussian noise channel, same waveform utilization
across adjacent channels, and a 10 $dB$ power imbalance (i.e., power
difference at the receiver). The results in Fig. \ref{fig:SER} reveals
that the proposed scheme performs better as expected due to lower
interference from the adjacent channel.

The higher spectral efficiency with the proposed scheme is realized
with increased computational complexity compared to conventional OFDM
systems. FFT/IFFT blocks require $Q\log(Q)$ multiplication operations
for the warped waveform, whereas $N\log(N)$ is required for OFDM.
Also, there are additional $Nl$ multiplication operations for the
pulse shaping, where $l$ represents the length of $g_{n}$. Finally,
the equalization consists of $N$ multiplication operations as similar
to OFDM. Another drawback of the warped waveform design is increased
PAPR. The increase is caused by expansion of the subcarriers and the
PAPR increases as the number of warped edge subcarriers increases.
However, our implementation shows that warping seven edge subcarriers
is sufficient for an efficient sidelobe suppression and only 0.5 $dB$
difference is measured at CCDF of $10^{-3}$ (measured as 11.3 $dB$).
Therefore, the PAPR of the warped waveform can be considered as comparable
with the PAPR of the conventional OFDM systems.

\section{Conclusions\label{sec:IV}}

The proposed warped waveform concept improves the flexibility of waveform
design and enhances the system performance with certain trade-offs.
It provides better frequency localization which is critical for asynchronous
transmission across adjacent sub-bands and harmony with other numerologies
in the network. Also, it has the flexibility to manage each side of
its spectrum differently by utilizing an asymmetric warping operator
(e.g., Eq. \ref{eq:Warp_Asym}). Hence, it prevents unnecessary OOBE
suppression that decreases the spectral efficiency when requirements
are different on each side of the band.

The results show the potential of such a flexible scheme and the design
will be optimized for a future study. For example, this letter demonstrates
the warped waveform concept with a multicarrier scheme. However, the
concept is applicable to single carrier schemes as well.

The forthcoming communication systems are advancing towards improved
flexibility in various aspects. Improved flexibility is crucial to
cater diverse service requirements, and definitely, the pulse shape
and time-frequency lattice should be a piece of the flexibility discussion
as well. The warped waveform concept contributes to this futuristic
goal and takes one step towards improved flexibility.

\bibliographystyle{IEEEtran}
\bibliography{IEEEabrv,WaveformRef2}

\end{document}